\documentclass{nature}
\usepackage{graphicx} 
\usepackage{graphicx}
\usepackage{makecell}
\usepackage{subcaption}
\usepackage{amsmath}
\usepackage{orcidlink}
\usepackage{lineno}
\PassOptionsToPackage{dvipsnames,usenames}{xcolor}
\usepackage{xspace}
\usepackage{url}
\usepackage[normalem, normalbf]{ulem}
\usepackage{xurl}

\makeatletter
\let\saved@includegraphics\includegraphics
\AtBeginDocument{\let\includegraphics\saved@includegraphics}

\renewenvironment*{figure}{\@float{figure}}{\end@float}
\makeatother

\newcommand{\unit}[1]{\ensuremath{\,\mathrm{#1}}}
\newcommand{\figref}[2]{Figure~\ref{#1}{#2}}
\newcommand\arcsec{\mbox{$^{\prime\prime}$}}

\newcommand{\Alfven}{Alfv\'{e}n\xspace}

\newcommand{\apj}{{Astrophys. J.}}
\newcommand{\apjl}{{Astrophys. J. Lett.}}
\newcommand{\apjs}{{Astrophys. J. Supp.}}

\newcommand{\aaps}{{Astron. Astrophys. Supp.}}

\newcommand{\aap}{{Astron. Astrophys.}}
\newcommand{\nat}{{Nature.}}
\newcommand{\pasj}{{Publ. Astron. Soc. Jpn }}

\newcommand{\ssr}{Space Science Reviews}
\newcommand{\solphys}{{\it Sol. Phys.}}
\newcommand{\grl}{{\it Geophys. Res. Lett}}

\usepackage{soul}

\newcommand{\xp}[1]{{\textcolor{black}{#1}}}

\title{Resolved magnetohydrodynamic wave lensing in the solar corona}

\author{Xinping Zhou\orcidlink{0000-0001-9374-4380}$^{1}$, Yuandeng Shen\orcidlink{0000-0001-9493-4418}$^{2}$\footnote[1]{email: ydshen@ynao.ac.cn}, Ding Yuan\orcidlink{0000-0002-9514-6402}$^{3,4}$\footnote[1]{email: yuanding@hit.edu.cn}, Rony Keppens\orcidlink{0000-0003-3544-2733}$^5$, Xiaozhou Zhao\orcidlink{0000-0003-2875-7366}$^5$, Libo Fu\orcidlink{0009-0003-8645-0992}$^{3,4}$, Zehao Tang\orcidlink{0000-0003-0880-9616}$^2$, Jiaoyang Wang\orcidlink{0009-0005-8494-4272}$^{3,4}$, Chengrui Zhou\orcidlink{0009-0005-5300-769X}$^2$}

\begin{document}

\maketitle

\begin{affiliations}
\item  College of Physics and Electronic Engineering, Sichuan Normal University, Chengdu 610068, People’s Republic of China
 \item Yunnan Observatories, Chinese Academy of Sciences, Kunming 650216, People’s Republic of China
 \item Shenzhen Key Laboratory of Numerical Prediction for Space Storm, Institute of Space Science and Applied Technology, Harbin Institute of Technology, Shenzhen, Guangdong, China
 \item Key Laboratory of Solar Activity and Space Weather, National Space Science Center, Chinese Academy of Sciences, Beijing, China
  \item Centre for mathematical Plasma Astrophysics, Department of Mathematics, KU Leuven, Celestijnenlaan 200B, B-3001 Leuven, Belgium 
\end{affiliations}

\noindent{\textcolor{black}{\bf Abstract}}\\
\\
\begin{abstract}
   \textcolor{black}{ Electromagnetic wave lensing, a common physical phenomenon recognized in visible light for centuries, finds extensive applications in manipulating light in optical systems such as telescopes and cameras. Magnetohydrodynamic wave is a common perturbation phenomenon in the corona. By using high spatio-temporal resolution observations from the Solar Dynamics Observatory, here, we report the observation of a magnetohydrodynamic wave lensing in the highly ionized and magnetized coronal plasma, where quasi-periodic wavefronts emanated from a flare converged at a specific point after traversing a coronal hole. The entire process resembles an electromagnetic wave lensing from the source to the focus. Meanwhile, the magnetohydrodynamic wave lensing is well reproduced through a magnetohydrodynamic numerical simulation with full spatio-temporal resolution. We further investigate potential applications for coronal seismology, as the lensing process encodes information on the \Alfven speed, in conjunction with favorable geometric and density variations.}
\end{abstract}

\noindent{\textcolor{black}{\bf Introduction}}\\
\\
Lensing of various wave phenomena, such as electromagnetic (EM) (light) or sound waves, is ubiquitous. Similar to a convex glass lens that focuses light, acoustic lenses can focus and manipulate acoustic waves and have a wide range of valuable applications, including biomedical engineering, imaging, cancer treatment, and surgery\cite{DICKSON,1998Ultrasound,2004High,2005High}. Acoustic lensing is responsible for the Rotunda effect \textcolor{black}{(sometimes called whispering galleries) \cite{raman1922whispering}}, where a conversation can be held on opposite sides of a round-shaped auditorium. An intuitive lensing effect in daily life also occurs in water waves: shallower water areas in the \xp{ocean} can act as lenses. Their amplitude and energy will be enhanced near the focal point when they pass through these areas\cite{Berry,2022arXiv220205926T}. Light maneuvering with optical lenses has been practiced for many centuries, which has evolved into a mature industry for lens manufacturing, such as cameras, telescopes, microscopes, and lasers. In general relativity, a ray of light follows the curvature of space-time, and its path would be bent around a massive object. This effect, known as gravitational lensing, could be used for detecting black holes and dark matter in the Universe\cite{1992grle.book.....S,2006Natur.439..437B}.

\textcolor{black}{Magnetohydrodynamic (MHD) waves play essential roles in various fundamental processes in the corona. For example, they might heat the corona plasma\cite{1961ApJ...134..347O,1999Sci...285..862N} and accelerate the solar wind\cite{2007Sci...318.1574D}. Moreover, through a technique known as coronal seismology\cite{1970PASJ...22..341U}, physical information carried by MHD waves can be utilized to decode coronal physical parameters (such as the strength of the magnetic field) that can not be measured directly by our current technology. Fast-mode magnetoacoustic waves are a type of MHD waves that perturb coronal magnetic field and plasma density simultaneously, enabling its detection in the extreme ultraviolet (EUV) bandpass. Although typical wave phenomena such as refraction, transmissions, and reflection in the corona have been reported in previous studies\cite{2013ApJ...773L..33S,2012ApJ...756..143O,2009ApJ...691L.123G}, the lensing effect of fast-mode MHD waves has not been directly observed.}

\textcolor{black}{Here, we present comprehensive \textcolor{black}{spatio-temporal} resolved observations of a globally propagating fast-mode MHD wave that exhibits an unambiguous lensing process upon its passage through a coronal hole (CH) that acts as a convex lens. The entire lensing process of the wave, including the transmission through the lens and the focusing at the focal point, is comprehensively captured by the Atmospheric Imaging Assembly (AIA)\cite{2012SoPh..275...17L} instrument onboard the Solar Dynamics Observatory (SDO)\cite{2012SoPh..275....3P} with full spatio-temporal resolution. Through numerical simulations, we not only reproduce the observed lensing effect but also conduct a comprehensive investigation of the essential physical parameters that govern the occurrence of lensing effects on MHD waves in structured plasma environments.}

\noindent{\textcolor{black}{\bf Results}}

\begin{figure}
 	\centering
 	\includegraphics[width=\textwidth]{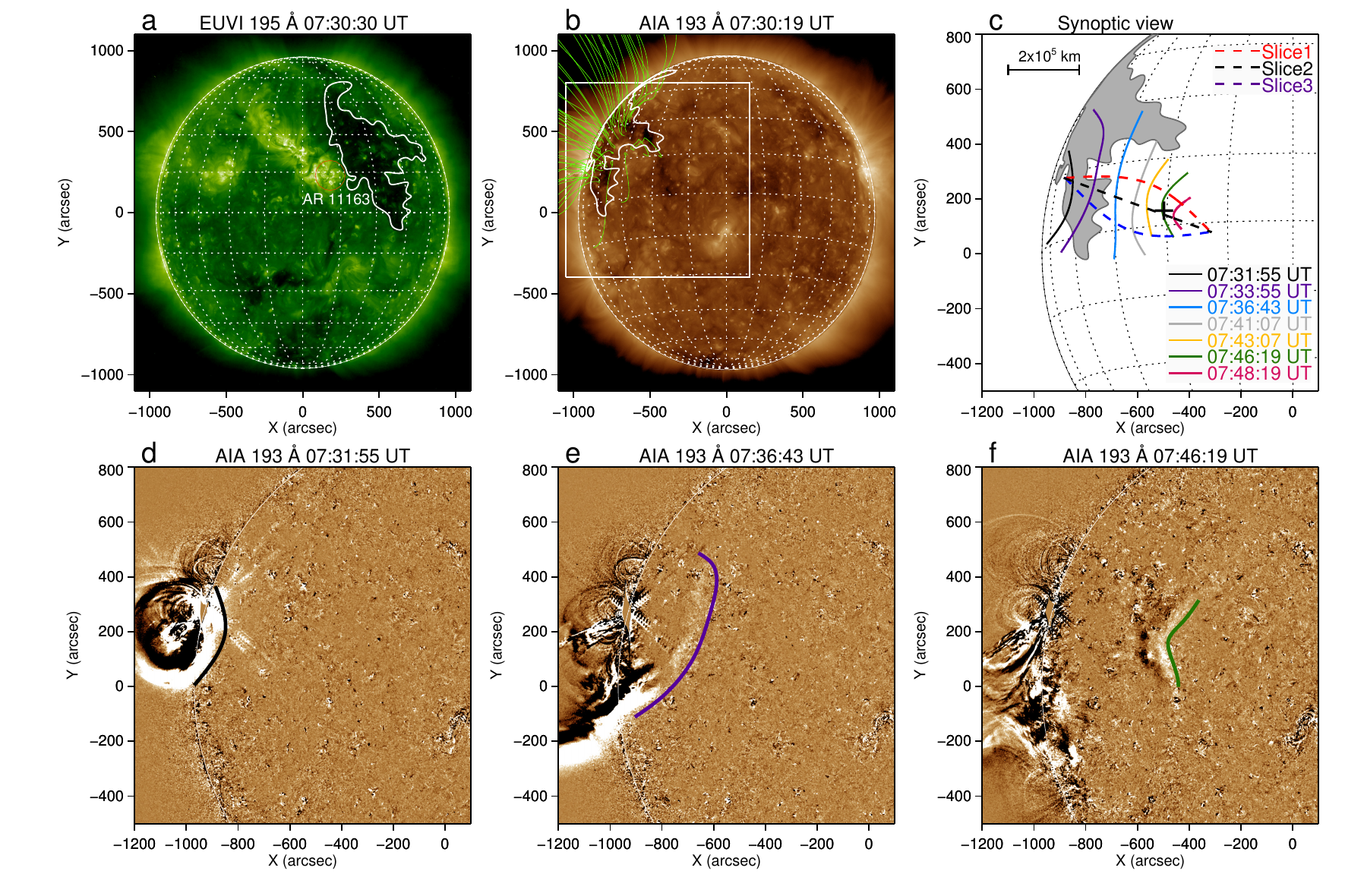}
 	\caption{\textcolor{black}{Overview of the solar coronal structure and MHD wave propagation and focusing process. (a)-(b) Combined views of STEREO-B/EUVI 195 \AA\ and SDO/AIA 193 \AA\ images (about 95 degrees apart) showing the full disk of the Sun at 07:30 UT on 2011 February 24. The red circle and white closed curve in (a) marked the active region AR11163 and the contours of the interest CH, respectively. The contour of the CH as deduced from STEREO-B/EUVI 195 \AA\ is also projected onto the SDO/AIA 193 \AA\ image. The open green curves in (b) show the extrapolated magnetic field line using the potential-field source-surface \textcolor{black}{(PFSS)} model\cite{1969SoPh....6..442S}, while the white box displays the \textcolor{black}{field} of view of (c)-(f). (c) Synoptic view showing the evolution of the MHD \xp{wavefront}. The CH is marked by a gray-shade, and the \xp{wavefront} is colored with time. The focal point at $(x,y)=[-500\arcsec, 150\arcsec]$ is labeled with a black plus sign. Three slits (Slice 1-3) are used to make time-distance plots as shown in \figref{fig:tdp}{} to track the evolution of wave. (d)-(f) SDO/AIA 193 \AA\ running-difference images showing the flaring core and the MHD wave evolution, where the evolving \xp{wavefront}\textcolor{black}{s are} labeled with colored curves. }(An animation Video I of this figure is available.) \label{fig:overview}}
 \end{figure}

On 2011 February 24, an energetic GOES M3.5 solar flare occurred in NOAA active region AR11163. \textcolor{black}{From the viewpoint of Solar TErrestrial RElations Observatory Behind (STEREO-B)\cite{2008SSRv..136....5K}, we can see a crescent-shaped CH adjacent to the active region, as shown in 195 \AA\ image captured by the Extreme UltraViolet Imager (EUVI)\cite{2004SPIE.5171..111W} onboard the STEREO (\figref{fig:overview}{(a)}).} The flare excited large-scale, quasi-periodic intensity perturbations propagating along the solar surface. These large-scale \textcolor{black}{perturbations}, a form of MHD waves\cite{2012ApJ...753...52L,2019ApJ...873...22S,2022ApJ...930L...5Z,2022SoPh..297...20S}, showed a chain of arc-shaped wavefronts, which remained approximately concentric with the flare center during the initial phase (\figref{fig:overview}{(d)}). \textcolor{black}{Subsquently,} the wave train propagated towards the center of the solar disk (\figref{fig:overview}(e)) and transmitted through a low latitude CH (the dark region in \figref{fig:overview}(a)-(b)). \textcolor{black}{Following,} the transmission, the wave train entered a quiet-Sun region with a speed of about $350\unit{km~s^{-1}}$. Such a speed is consistent with the typical \Alfven speed of the quiet corona\cite{2013ApJ...776...58N}. Interestingly, as the wavefronts propagated westward after crossing the CH's far-side boundary, the original arc-shaped \xp{wavefronts} gradually became anti-arc-shaped and converged towards the focal point far from the CH (\figref{fig:overview}{(f)}).\textcolor{black}{ The complete evolution of the MHD wave is display in \figref{fig:overview}{(c)}. More details can be found in Supplementary Video I. } These observational characteristics strongly suggest that the observed wave train is a propagating quasi-periodic fast-mode MHD wave in nature, and its physical property is similar to the globally propagating EUV waves \cite{1968SoPh....4...30U,1998GeoRL..25.2465T,2000ApJ...543L..89W,2002ApJ...572L..99C}. In addition, the whole evolution process of the wave train from its source to the focusing resembles an EM wave passing through a converging lens; or mimics acoustic wave refocusing, but this time dictated by geometric variation of the \Alfven speed. \textcolor{black}{ In the present case, the flare center, the CH, and the final focal point make up the key components of the MHD lensing system.} The lensing effect of the wave train could be detected clearly in \textcolor{black}{ AIA 211 \AA\ (2.0 MK), 193 \AA\ (1.6 MK), and 171 \AA\ (0.6 MK)} channels. Here, we only present the observations of the AIA 193 \AA\ channel.

 A CH is characterized by dark regions in X-ray and EUV images, where a single magnetic polarity dominates\cite{1996Sci...271..464W}. At the same time, the densities and temperatures are lower than in the CH surroundings\cite{1972ApJ...176..511M,2004A&A...416..749A,Cranmer2009}. Therefore, a CH is often a region of high \Alfven speed with respect to the quiet-Sun. Theoretical investigation suggests that large-scale coronal MHD waves will be refracted away from regions of high \Alfven speed towards the low \Alfven speed areas. Meanwhile, as a linear fast-mode MHD wave enters the low \Alfven speed region, it may steepen into a shock wave due to refraction and the deceleration of the wave\cite{1968SoPh....4...30U,1974SoPh...39..431U}. For an optical lensing system, the light speed is usually reduced significantly inside the lens with respect to that outside the lens. However, for the MHD lensing system presented here, the fast-mode MHD wave propagated faster inside the CH than in the quiet-Sun region\textcolor{black}{;} it also showed an obvious focusing effect after its transmission through the CH \textcolor{black}{due to the special shape of the CH that acts as the MHD lens}. \textcolor{black}{The behavior of the observed wave train is hence consistent with theoretical prediction\cite{1974SoPh...39..431U}.} The particular crescent geometry and enhanced \Alfven speed of the CH \textcolor{black}{are} important for the occurrence of the observed lensing effect.

\begin{figure}
 	\centering
 	\includegraphics[width=1\textwidth]{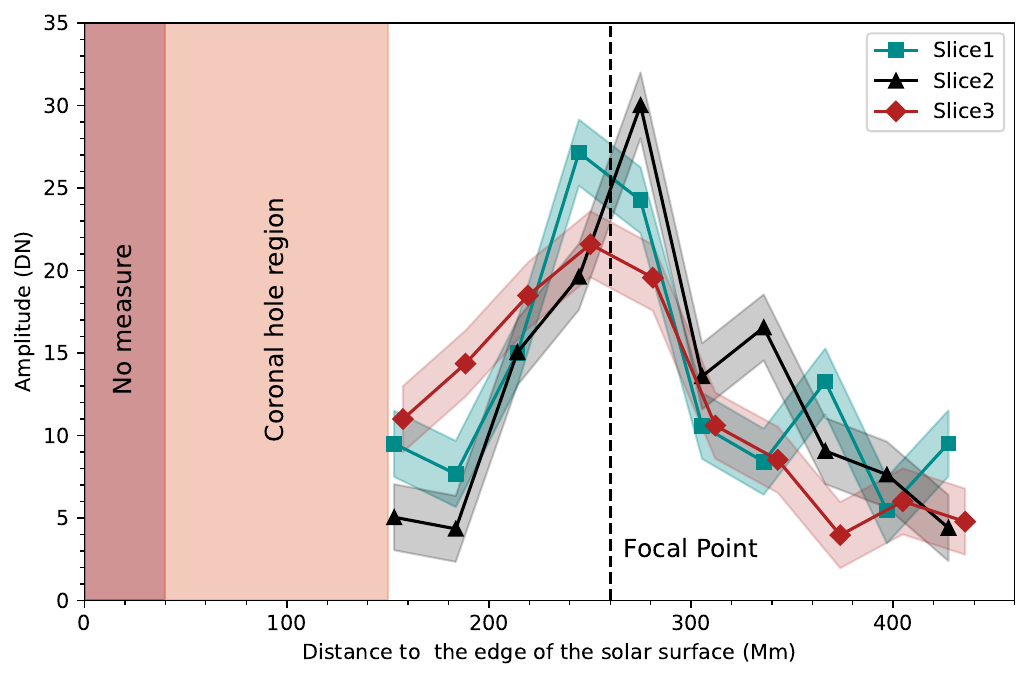}
 	\caption{\textcolor{black}{Intensity amplitude evolution of the MHD wave along three slits indicated by Slice 1, Slice 2, and Slice 3 in \figref{fig:overview} (c). The data sampling locations correspond to the positions marked by the white horizontal lines p1-p10 in the time-distance plots of \figref{fig:tdp}. The colored shades give the {errors obtained from the calculation method \textcolor{black}{provided} by Yuan et al\cite{2012A&A...543A...9Y} for wave amplitude in units of data number (DN)}. The location of the CH was depicted in light brown color, while the brown area represents regions that were not measured due to overexposure. The vertical dashed line marks the position of the focal point.} \label{fig:amplitude}} 
 \end{figure}

To quantify the MHD lensing effect, we measure the intensity amplitude and energy flux density of the wave train along its propagation path. Although the STEREO-B had an advantageous viewing angle to observe the eruption source region, its spatial and temporal resolutions are too low to resolve the lensing process. Luckily, the SDO/AIA images are sufficient to determine the detailed evolution process of the wave, though its viewing angle was not ideal for measuring the wave signal before and during the transmission of the wave train through the CH. As shown in \figref{fig:overview}{(b)}, the eruption source region was on the backside of the Sun, and the wave signal inside the CH was too weak to measure. Thus, we only measure the intensity amplitude variation after the wave train passed through the CH using the AIA images, i.e., during the focusing phase. \figref{fig:amplitude}{} shows the intensity amplitude variations along three selected paths (\figref{fig:overview}{(c)}). As the wave approached the focal point, the intensity amplitude increases by about 2 to 6 times along all three paths. The intensity amplitudes reached their peaks around the focal point, and then they started dropping off quickly to the level before the focusing. We calculate the energy flux density of the wave to estimate the energy-focusing effect (see Methods section). The results indicate that energy flux densities of the wave were respectively about $5.2\times10^4\unit{ erg\cdot cm^{-2}\cdot s^{-1}}$ and $3.5\times10^5\unit{ erg\cdot cm^{-2}\cdot s^{-1}}$ before the focusing and at the focal point, indicating that the wave energy flux density at the focal point increased about seven times \textcolor{black}{compared to the pre-focusing level. Therefore\textcolor{black}{,} the results suggest} that an MHD lens can effectively focus wave energy on the focal point. Since magnetized plasma and magnetic field pervade the whole Universe and the ideal MHD description is scale-invariant, such MHD lensing is expected to be duplicated in magnetized planetary, stellar, and galactic counterparts and may play a significant role in the convergence of MHD wave energy.

As mentioned above, the observation suggests that a particular structuring of magnetized plasma could act as an MHD lens that maneuvers an MHD wave and the associated energy flux. We model this scenario in a 2.5-dimensional magnetized plasma to further demonstrate the MHD lensing effect. The essential aspects of the lensing effect of a fast-mode MHD wave can be investigated in a simple idealized setup, where the influence of the model parameters can be shown clearly. This part is done in \textcolor{black}{Methods section}, where the numerical approach and setup details are also given. Here, we discuss a more realistic scenario that uses the observed geometric shape of the CH. 

\begin{figure}
 	\centering
 	\includegraphics[width=1\textwidth]{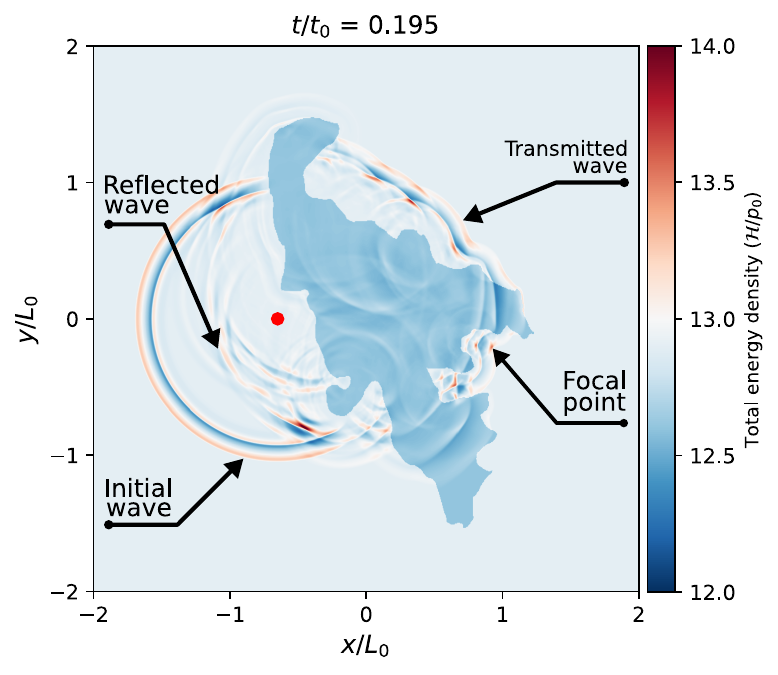}
	 	\caption{ \textcolor{black}{{\bf Numerical simulation of the MHD lensing process at $t/t_0=0.185$ based on the observed geometric shape of the CH}. The simulation domain includes a background coronal plasma (light blue) and the CH that acts as an MHD lens ({blue}), as shown in the {total energy density ($\mathcal{H}/p_0$)}. A single-period fast-mode MHD wave is launched to the left of the CH, which propagates with a circular wavefront initially. After being reflected/transmitted, the initial circular wavefront presents a complex pattern that senses the geometric shape at the CH edges. The initial {wavefront}, the reflected {wavefront} by the CH, the transmitted wave, and the focal point are annotated by black arrows. The values of normalization factors $p_0$, $t_0$, $L_0$ for the total energy density ($\mathcal{H}$), time ($t$) and length ($x,y,z$) are listed in Table~\ref{tab:norm}. The red dot represents the location of the wave source, and it retains the same meaning in the subsequent Figures.} (An animation Video II of this figure is available.)\label{fig3}
 	}
 \end{figure}

\textcolor{black}{ As shown in \figref{fig3}{}, t}he simulation ignores the solar surface's curvature and takes the CH's shape from the STEREO-B/EUVI 195 \AA\ image as shown in \figref{fig:overview}{\textcolor{black}{(a)}}. The location of active region AR11163 relative to the CH is also determined by this image. We launch a point-source induced fast-mode MHD wave from the center of the active region, which shows as a circular \xp{wavefront} and travels at the local \Alfven speed away from the source region. The rightward edge of the induced \xp{wavefront} gets reflected and transmitted at the left edge of the CH. \textcolor{black}{\textcolor{black}{Eventually}, some of the transmitted waves converge to a particular point. More details can be \textcolor{black}{found} in Supplementary Video II.} Transmitted wave patterns traverse the CH at the locally higher \Alfven speed, in direct correspondence to the adopted density (and temperature) contrast used in the model. The background magnetic field was set to be orthogonal to the simulation plane (i.e., a vertical, open field is assumed, as appropriate for the CH region). The sum of gas and magnetic pressures in the CH region was set in balance with the surrounding coronal plasma, where input parameters control the external plasma beta (here $\beta=0.05$), and a prescribed density and temperature contrast reflects the observed lower $T$ and $\rho$ conditions in the CH. After using reasonable density and temperature ratios, wave amplitudes, and source duration, the CH shape dictates the wave behavior throughout the simulated coronal region. We find that the complex shape of the CH causes a single circular wavefront to become reflected/transmitted in an intricate pattern of constructively and destructively interfering waves that sense the changes in background quantities at the CH edges. After the first transmitted wave has wholly traversed the CH, the concave part found at the opposing side of the wave source region causes an intricate pattern of constructively interfering waves in a location similar to the observations that find the lensing to occur. We can then isolate the main ingredients of an \Alfven-speed-lensing in a different idealized setup by simplifying the CH shape to be composed of a straight left edge with a concave parabolic right edge. This different idealized model is then parametrically explored in \textcolor{black}{Methods section} (this has the advantage that the complex wave interference patterns due to the complex CH shape boundary are omitted). There, we find that the density ratio determines the precise shape of the lensing at the focal point, and the overall plasma beta sets the time of traversal from source to focus. In the idealized setup, after passing the artificial MHD lens (or CH), the amplitude increases as it approaches the focal point and then becomes weaker as it moves away from the focal point. The simulation's wave propagation and focusing phase are consistent with our observation, implying that observing a lensing event allows us to deduce local \Alfven speed, density variations, and plasma beta conditions. At the same time, the needed geometric information can be directly seen in EUV images. This is of immediate use for coronal seismology.

\noindent{\bf Discussion}

This paper reports an observation of the MHD lensing effect with full spatial and temporal resolution. A quasi-periodic fast-mode MHD magnetoacoustic wave was excited by a solar flare and transmitted through a structured CH with low density. Due to the particular shape and the higher \Alfven speed of the CH, it acted as an MHD converging lens and focused the wave energy towards a small region beyond the CH, although the focussing is not total. Thanks to the full disk and high spatio-temporal resolution observations taken by SDO/AIA, we successfully observed the dynamical lensing effect of MHD waves in solar, magnetized plasma conditions. We demonstrated that the lensing of a fast-mode magnetoacoustic wave conducted by a CH follows our numerical experiment, where the main parameters influencing the lensing process can be identified in a simple idealized setup. We note that deflection, reflection, and transmission effects of globally propagating EUV waves have been observed and confirmed by many numerical simulations during their interaction with coronal structures such as active regions and CHs\cite{2000ApJ...543L..89W, 2002ApJ...574..440O, 2006ApJ...647.1466V, 2009ApJ...691L.123G, 2012ApJ...756..143O, 2013ApJ...773L..33S, 2018A&A...614A.139A, 2021A&A...651A..67P, 2022A&A...659A.164Z}. However, none of these works considered the influence of the shape of CHs on wave propagation. In the present case, the appearance of the lensing effect is believed to be due to the sharp gradients of the temperature, background plasma density, and magnetic field strength at the CH boundary, as well as the particular shape of the CH, which could be explained using the method of geometrical acoustics\cite{1968SoPh....4...30U}. Another focus of coronal fast-mode MHD waves appears at magnetic null points that commonly exist in breakout \cite{1999ApJ...510..485A,2012ApJ...750...12S} and fan-spine\cite{2009ApJ...700..559M,2019ApJ...885L..11S} magnetic systems. Since the magnetic field at a coronal null point approaches zero, the low \Alfven speed property around the null point will cause a fast-mode MHD wave to focus towards and wrap around the null point when the wave approaches the null\cite{2011SSRv..158..205M}. This convergence of an MHD wave at a coronal null point differs from the focusing of an MHD wave at a non-zero \Alfven speed point, as is happening in the MHD lensing effect discussed here, though they \textcolor{black}{are both} due to refraction effects. In the first case, the wave and its energy accumulate at the zero point, resulting in a convergence effect due to the inability of the wave to pass through the magnetic zero point\cite{2008SoPh..251..563M}. The focusing effect reported in the present paper is a manifestation of magnetoacoustic focusing in the solar coronal plasma environment and can be explained in terms of classical geometric acoustics\cite{1968SoPh....4...30U}. 

The MHD lensing effect documented in this study must have counterparts across various length scales since the MHD description of plasma dynamics is scale invariant\cite{Hansbook3}. A localized and specially shaped MHD lensing setup could focus MHD wave energy to a specific destination. This process can be reproduced and experimented with in laboratory plasmas. The MHD lens could be designed to either converge or diverge MHD wave energy. For example, the magnetosphere of the Earth diverts solar storms from the dayside to the magnetotail. Therefore, it can be viewed as a natural diverging MHD lens that effectively protects life on the Earth\cite{1993GeoRL..20.2809P}. A converging MHD lens is observed and modeled, showing clear diagnostic information on the background plasma. We illustrated how the focus relies on low plasma beta conditions and specific shapes of CHs and encodes their density contrast and \Alfven speed variations. Since fast-mode MHD waves can be easily excited in space and laboratory plasma, the MHD lensing effect could be used for diagnosing various plasma properties, as initially proposed for solar coronal Moreton waves\cite{1968SoPh....4...30U}.

\clearpage

\vspace{1.5cm}
\noindent{\bf Methods}\\
\\
\noindent{\bf Observations and Data Processing}\\
This study used imaging observations provided by AIA. The AIA takes full-disk corona images at seven EUV wavelength bands that are sensitive to plasma emissions with a wide temperature coverage ranging from about 500,000 K to over 20,000,000 K. Its time resolution is 12 seconds, and each pixel corresponds to an angular width of $0.6$ arc second or about 420 km on the Sun. In this study, we focus on the 193 \AA{} bandpass (Fe XII line) that is mainly from plasma emission at about 1,000,000 K. We also used images taken by the EUVI onboard the STEREO. EUVI observes the corona with an angle of \textcolor{black}{about $95^\circ$} apart from AIA with respect to the center of the Sun. In the EUVI's viewpoint, AR11163 and the CH were observed with negligible projection effects (\figref{fig:overview}{(a)-(b))}. Both AIA and EUVI images were calibrated with standard routines in the Solar Software (SSW). This process calibrates the images by removing the CCD bias and dark frames, correcting for the flat field, and then normalizing them with the exposure time.

\begin{figure}
 	\centering
 	\includegraphics[width=1\textwidth]{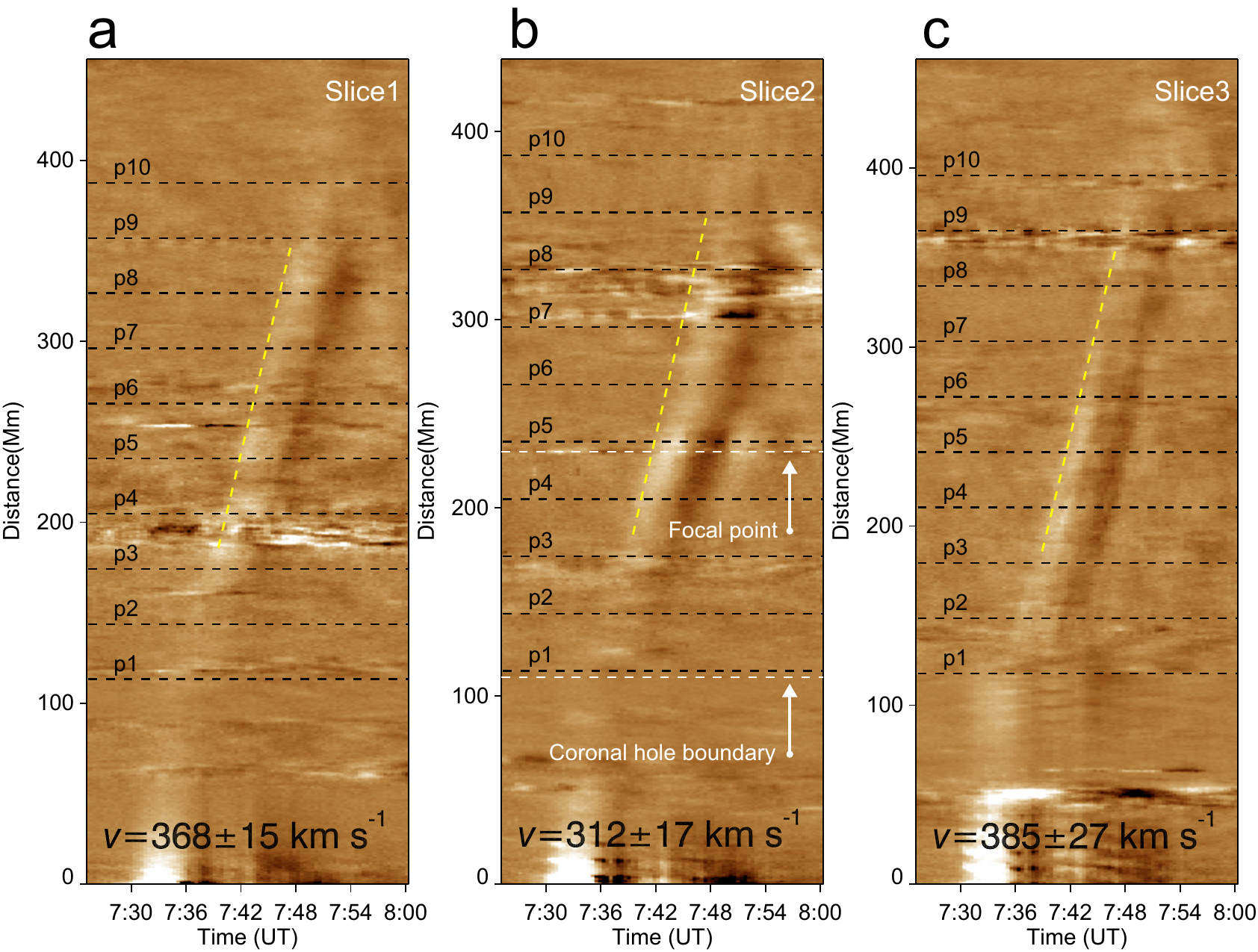}
 	\caption{\textcolor{black}{{\bf Dynamics of the observed MHD wave.} (a)-(c) Time-distance plots made from the AIA 193 \AA\ base difference images along the three paths Slice 1, Slice 2, and Slice 3 shown in \figref{fig:overview}{(c)}. The black horizontal lines marked with p1--p10 in each panel indicate the positions used to extract the signal for analyzing wavefront amplitudes in \figref{fig:amplitude}{}. These positions are selected at an equal interval \textcolor{black}{of} 30 Mm and originate about 110 Mm from the starting point of the slices. The white horizontal lines labeled with white arrows mark the focal point and the far-side boundary of the coronal hole. In contrast, the oblique dotted yellow lines depict the linear fitting results for estimating the speeds of the wave. The speeds of the waves along different slices are listed in the corresponding panel. {The error is estimated by making the fit ten times.}} }
 	\label{fig:tdp}
 \end{figure}

\noindent{\bf Kinematic Study and Energy Estimation}\\
In order to highlight the \xp{wavefronts}, we calculated the forward running difference of the image sequence in the time dimension, as shown in \figref{fig:overview}{(d)-(f)}. The \xp{wavefronts} were tracked manually to measure wave parameters, as illustrated in \figref{fig:overview}{(c)}. The emission intensities along these slits were stacked in time to form time-distance plots. \figref{fig:tdp}{} shows the \xp{wavefronts} propagation in the form of base-difference time-distance maps, in which the slopes of the ridges represent the speeds. In \figref{fig:tdp}{}, we can see that the speed of the wave along Slice 2 (about $312\unit{km~s^{-1}}$) is smaller than Slice 1 (about $368\unit{km~s^{-1}}$) and Slice 3 (about $385\unit{km~ s^{-1}}$) after its passage through the CH, since the paths of the latter two are longer. Additionally, the \xp{wavefront} showed a visible intensity enhancement at the focal point as shown in \figref{fig:tdp}{(b)}. The average phase speed is $350\unit{km~s^{-1}}$ (with projection effect), which is in agreement with the typical fast-mode MHD wave speed in the solar corona \cite{2013ApJ...776...58N}. The black horizontal lines in \figref{fig:tdp}{} labeled p1-p10 indicate the positions for extracting the amplitudes (emission intensities) of the wave. After obtaining the value of the wave crest at each position, we get the amplitude variations along the three paths, as shown in \figref{fig:amplitude}{}. The uncertainties on the wave amplitude were estimated mainly by accounting for the photon noise, as suggested by Yuan et al.\cite{2012A&A...543A...9Y}

Considering that the group velocity of the wave is more difficult to obtain, we do not use the equation for calculating the energy flow using the group velocity suggested elsewhere\cite{2014ApJ...795...18V} but employed the following equation for estimating the energy flow using the phase velocity:
\begin{equation}
	\mathcal{F} =\frac{1}{2}\rho (\delta v)^2v_\mathrm{ph}, \label{eq:flux}
\end{equation}
where $\rho$ is the plasma density, $\delta v$ is the perturbation speed of the local plasma, and $v_{\mathrm{ph}}$ is the phase speed of the wave\cite{2004psci.book.....A,2011ApJ...736L..13L}. The relative perturbation speed could be estimated by $\delta v/v_{\mathrm{ph}} \geq \delta \rho/\rho = \delta I /(2I)$, where $\delta \rho$ and $ \delta I$ are the perturbations to density and emission intensity. We assumed that the emission intensity of the AIA 193 \AA{} bandpass is proportional to the square of density (and hence the square of electron number density\cite{1997A&AS..125..149D,2021ApJ...909...38D}), $I \propto \rho^2 \propto n_\mathrm{e}^2 $. Then Equation~\ref{eq:flux} could be rearranged for practical use  as done in Ref\cite{2018ApJ...860...54O} 
\begin{equation}
	\mathcal{F} \geq \left(\frac{n_\mathrm{e}}{10^8\unit{cm^{-3}}}\right)\left(\frac{\delta I/I}{1\%}\right)^2\left(\frac{v_\mathrm{ph}}{1000\unit{km\cdot s^{-1}}}\right)^3\left[2.5\times10^3\unit{ erg\cdot cm^{-2}\cdot s^{-1}}\right].
\end{equation}
$\delta I/I$ was measured to be about 12\% at the right boundary of the CH, while it was about 31\% at the focal point. If we use the average phase speed of $v_\mathrm{ph}=350\unit{km~s^{-1}}$ and number density $n_\mathrm{e}=3.4\times10^8\unit{ cm^{-{3}}}$  (an empirical value\cite{2011ApJ...730..122W}), then we obtain that the energy flux of the transmitted wave through the CH was about $5.2\times10^4\unit{ erg\cdot cm^{-2}\cdot s^{-1}}$, while the energy flux at the focal point was about $3.5\times10^5\unit{erg\cdot cm^{-2}\cdot s^{-1}}$. This means that the energy flux of the wave increased by about 6.7 times after the focusing.

\noindent {\bf Numerical model and analysis}\\
To model a fully ionized and magnetized plasma in macroscopic scales, we used the Newtonian MHD equations, a scale-invariant model for plasma dynamics\cite{Hansbook3}. They combine the pre-Maxwell equations of electromagnetism and the equations of fluid dynamics as follows,
\begin{align}
	\frac{\partial\rho}{\partial t}+\nabla\cdot \left(\rho\mathbf{v}\right) &=0, \label{eq:1} \\
	\frac{\partial\left(\rho\mathbf{v}\right)}{\partial t}+\nabla\cdot \left[\rho\mathbf{v}\mathbf{v}+\left(p+\frac{\mathbf{B}^{2}}{8\pi}\right)\mathbf{I}-\frac{\mathbf{B}\mathbf{B}}{4\pi}\right] &=  \mathbf{0},\label{eq:2} \\
	\frac{\partial \mathcal{H}}{\partial t}+\nabla\cdot \left[\left(\mathcal{H}-\frac{\mathbf{B}^{2}}{8\pi}+p\right)\mathbf{v}-\frac{(\mathbf{v}\times\mathbf{B})\times\mathbf{B}}{4\pi}\right] &= 0,\label{eq:3} \\
	\frac{\partial\mathbf{B}}{\partial t}+\nabla\cdot \left(\mathbf{v}\mathbf{B}-\mathbf{B}\mathbf{v}  \right)& = \mathbf{0} ,\label{eq:4} \\
	\nabla \cdot \mathbf{B}=0,  \label{eq:5} 
\end{align}
where $\rho$, $p$, $\mathbf{v}$, and $\mathbf{B}$ are the mass density, the gas pressure, the fluid velocity, and the magnetic field, respectively. Here $\mathcal{H}=[\rho\varepsilon+\rho\mathbf{v}^{2}/2+\mathbf{B}^{2}/8\pi]$ is the total energy density, where $\varepsilon=p/(\gamma \rho-\rho)$ is the internal energy per unit mass. We adopted $\gamma=5/3$ as the adiabatic index. This set of MHD equations is closed by the equation of state
$p=\rho k_{\mathrm{B}}T/\mu m_{\mathrm{H}}$, where $m_{\mathrm{H}}$ is the hydrogen atom mass, and the mean molecular weight is taken as $\mu = 1.4/2.3$ for a fully ionized plasma with a $10:1$ abundance of hydrogen and helium.

We used the open-source MPI-AMRVAC code\cite{2012JCoPh.231..718K,2014ApJS..214....4P,2021amrvac,2023amrvac} to solve the MHD equations in a conservative form with a finite-volume scheme. In our setup, we used an HLLD solver with a Koren-type limiter for reconstruction and a three-step Runge-Kutta method in the time integration. Equations~\ref{eq:1}-\ref{eq:5} were solved in a dimensionless form. The normalization uses the reference quantities $L_0$, $\rho_0$, and $B_0$ etc., as given in Table~\ref{tab:norm}.

We modeled a CH surrounded by a uniform and static coronal plasma. The essential features of fast-mode lensing can be investigated both in a simple idealized setup and a more realistic scenario that uses the observed geometric shape of the CH. The latter is already shown in \textcolor{black}{ \figref{fig3}{}}. Here, we describe the numerical setups and present the simulation results of the idealized setup in \textcolor{black}{detail}. 

\begin{figure}
	\centering\textbf{}
	\includegraphics[width=1\textwidth]{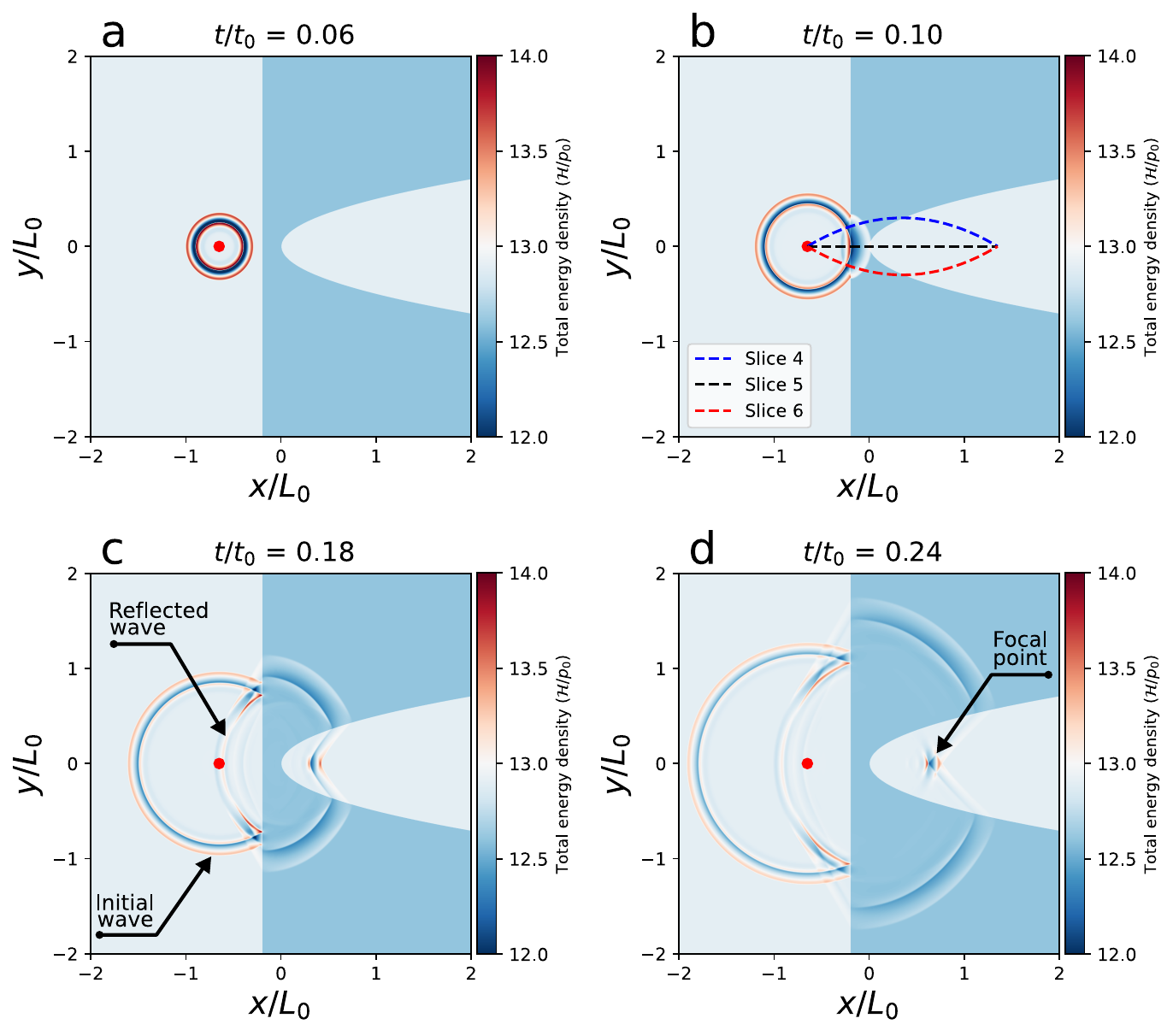}
	\caption{ \textcolor{black}{{\bf Evolution of perturbation to the \xp{total energy density ($\mathcal{H}/p_0$)} during \textcolor{black}{the} MHD lensing process in an idealized numerical simulation.} (a) Numerical set-up and MHD wave excitation. The simulation domain includes a background coronal plasma (light blue) and an artificial MHD lens (blue) enclosed by a straight line and a parabolic curve. A fast-mode MHD wave train launched to the left of the MHD lens propagates outwards in the form of concentric rings. (b) Initial phase of MHD wave propagation and its interaction with the MHD lens. (c) and (d) MHD wave focusing process. The black arrows point out the initial wave, the reflected wave and the location of the focus. Three paths (Slice 4, Slice 5\textcolor{black}{,} and Slice 6) in (b) are used for making the time-distance plots as shown in \figref{ap2fig2}. The values of normalization factors $p_0$, $t_0$, $L_0$ for the total energy density ($\mathcal{H}$), time ($t$) and length ($x,y,z$) are listed in Table~\ref{tab:norm}. 
		(An animation Video III of this figure is available.)}}
	\label{ap2fig1}
\end{figure}

For the idealized setup, the simulation domain includes a low-density (high Alfv\'en speed) region to act as an artificial MHD lens. This region is equivalent to a CH, and it is bounded by a straight line $x/L_{0}=-0.2$ at the left edge and a parabolic curve $x/L_{0}=4y^{2}$ (see \figref{ap2fig1}) at the right edge. The background coronal plasma has a density of $\rho_\mathrm{bg}/\rho_{0}=1$ and a pressure of $p_\mathrm{bg}/p_{0}=1/\gamma$. 
The CH was set to have a cooler temperature and lower density than the background. The density and temperature ratios between the CH and background plasma are $R_{\rho}=0.25$ and $R_\mathrm{T}=0.1$, respectively.
The magnetic field was set normal to the 2D simulation plane; therefore, it only has a $z$-component initially. The background magnetic field strength was set to $B_\mathrm{bg}/B_{0}=\sqrt{2p_\mathrm{bg}/( p_{0}\beta_\mathrm{bg})}$, where $\beta_\mathrm{bg}=0.05$ is the plasma beta for the background coronal plasma, i.e. the ratio of the gas pressure to the magnetic pressure. The magnetic field of the CH was calculated by setting the total plasma pressure ($p_\mathrm{tot}=p+\mathbf{B}^2/8\pi$) in equilibrium over the simulation domain. 
\textbf{}

This simulation was done in a two-dimensional plasma region of $-2\leq x/L_{0}\leq 2$ and $-2\leq y/L_{0} \leq 2$, where $L_{0}$ is our normalization factor for length, see Table~\ref{tab:norm}. The region has a base computational grid of $100\times100$ cells,  but an adaptive mesh refinement (AMR) was set up to allow for five levels. Therefore, the finest grid reaches a physical cell size of $700\,\mathrm{km}$. At the lateral boundaries, the {\it z}-component of the velocity $v_{z}$ and the {\it x} and {\it y}-components of the magnetic field $B_{x}$ and $B_{y}$ are fixed to zero while a continuous extrapolation applies to other variables.

A periodic fast-mode MHD wave is excited by setting a perturbation source located at $(x_\mathrm{s}/L_0,y_\mathrm{s}/L_0)=(-0.65,0)$ in the following form,
\begin{align}
	\rho_1/\rho_{0} &=  {\Gamma_{v}\alpha_f \omega_f /(k_{w}v_{0})}, \\
	p_1/p_{0}      &=   { \Gamma_{v}\alpha_f \omega_f /(k_{w}v_{0})}, \\
	v_{x1}/v_{0}   &=  \Gamma_{v}\cos\theta  ,\\
	v_{y1}/v_{0}   &= \Gamma_{v}\sin\theta,\\
	B_{z1}/B_{0}   &=   { \Gamma_{v}   B_{\omega}k_{w}v_{0}/ \omega_f  } , 
\end{align}
where $\Gamma_{v}=P_{b}\omega_f\cos(\omega_f t)$, $\omega_f=k_{w}v_{0}\sqrt{B_{\omega}^2+C_{\omega}^2}$, $B_{\omega}=B_\mathrm{bg}/B_{0}$,  $C_{\omega}=1$, $k_{w}=2\pi/\lambda_w$, $\lambda_w/L_{0}=0.1$, and $P_{b}=0.5$. Furthermore, we have $K_{b}=k_{w}B_{\omega}v_{0}/\omega_f$, $\alpha_f=1-K_{b}^2 $.
This, in essence, corresponds to the exact analytic eigenfunction variation for a fast (linear) magnetoacoustic wave\cite{Hansbook3} with frequency $\omega_f$. Geometric factors are centered on the source location as in $\cos\theta = (x-x_{s})/r_{0}$, $\sin\theta = (y-y_{s})/r_{0}$, where $r_{0} = \sqrt{(x-x_{s})^2 + (y-y_{s})^2}$ and $r_{0}<0.05L_{0}$.

\begin{figure}
	\centering\textbf{}
	\includegraphics[width=.5\textwidth]{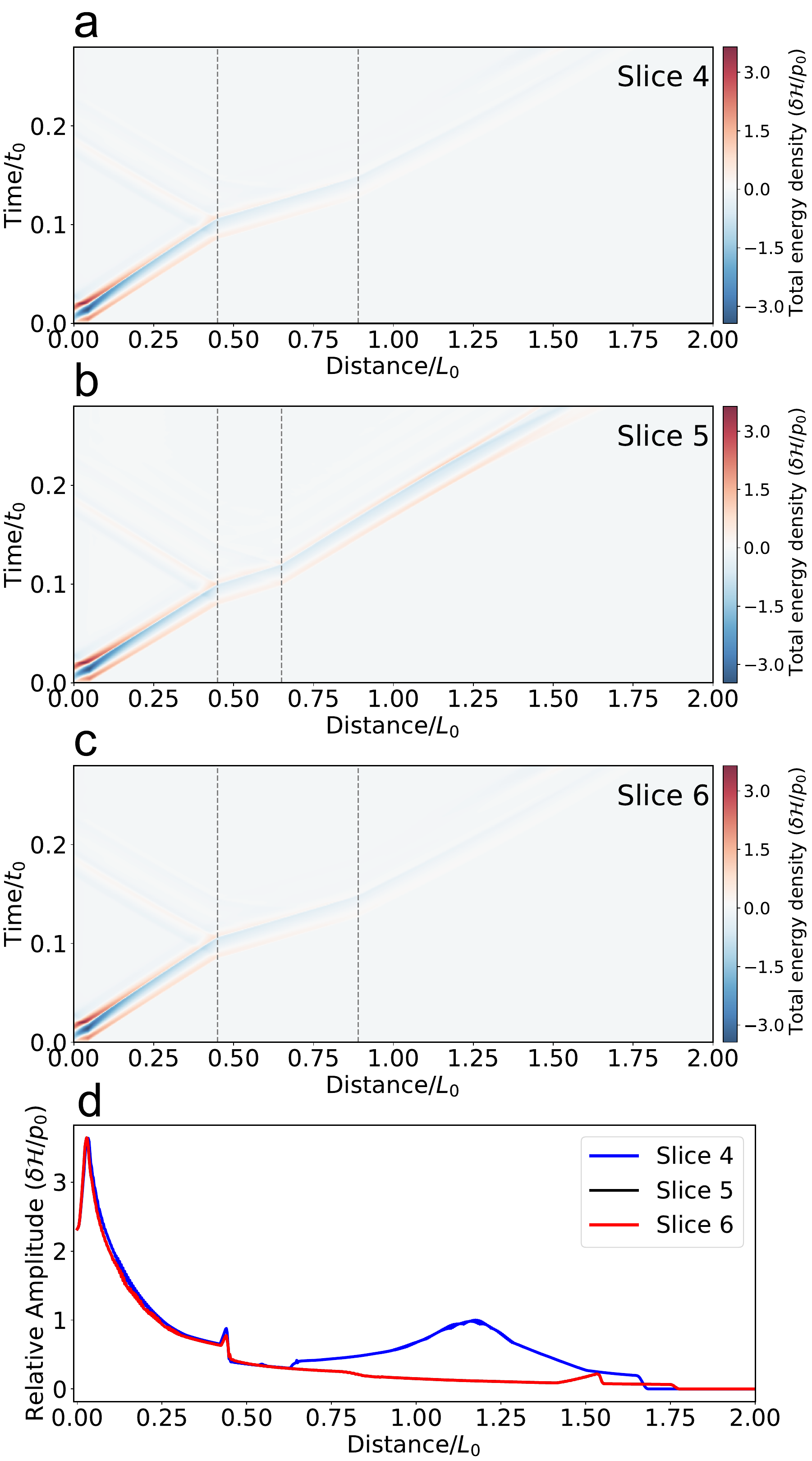}
	\caption{\textcolor{black}{{\bf Dynamics of the simulated MHD wave and its amplitude variations along three selected paths.} (a)-(c) Time-distance plots to show the total energy density $\delta \mathcal{H}/p_0$ along the paths as shown in \figref{ap2fig1} (b), where $\delta \mathcal{H}$ is the difference between total energy density $\mathcal{H}$ and the background total energy density $\mathcal{H}_0$. The vertical dotted lines indicate the CH boundaries in the numerical simulation. (d) The amplitude variation of the wavefront along the selected paths. As the upper/lower paths shown in \figref{ap2fig1}{(b)} are symmetric to each other, the measured values are identical. The values of normalization factors $p_0$, $t_0$, and $L_0$ for the total energy density ($\mathcal{H}$), time ($t$) and length ($x,y,z$) are listed in Table~\ref{tab:norm}}.}
 \label{ap2fig2}
\end{figure}

The \xp{wavefronts} propagate outward radially in the form of concentric rings; partial reflection and transmission occur at the left boundary of the MHD lens. When the MHD wave passes throingugh the MHD lens, it starts to bend towards a focal point. The convergence process is similar to the observation, i.e., the \textcolor{black}{focusing} is not total, which is caused by the shape of the CH. \textcolor{black}{More details can be found in Supplementary Video III.} The focal length is clearly related to the geometry and structuring of the MHD lens. \textcolor{black}{Along the selected path, we first extract the slices from the total energy density $\delta\mathcal{H}/p_0$ images (here the $\delta \mathcal{H}$ is the difference between total energy density $\mathcal{H}$ and the background total energy density $\mathcal{H}_0$ to remove the background and retain the wavefront signal.), and subsequently stack these slices according to the time series to obtain the time-distance plots, as shown in \figref{ap2fig2}{(a), (b), and (c)}.} \textcolor{black}{Based on these obtained time-distance plots of the wavefront evolution, w}e then calculated the wave amplitude variation following three paths \textcolor{black}{(Slice 4, Slice 5 and Slice 6 in \figref{ap2fig1}{(b)})} starting from the source to the focal point. \textcolor{black}{As shown in \figref{ap2fig2}{(d)}, t}he wave amplitude decreases as the wave departs from the source. After passing the artificial MHD lens (or CH), the amplitude increases as it approaches the focal point and then becomes weaker as it moves away from the focal point. The wave transmission and focusing phase in the simulation are consistent with our observation.

\begin{figure}
	\centering
	\includegraphics[width=1\textwidth]{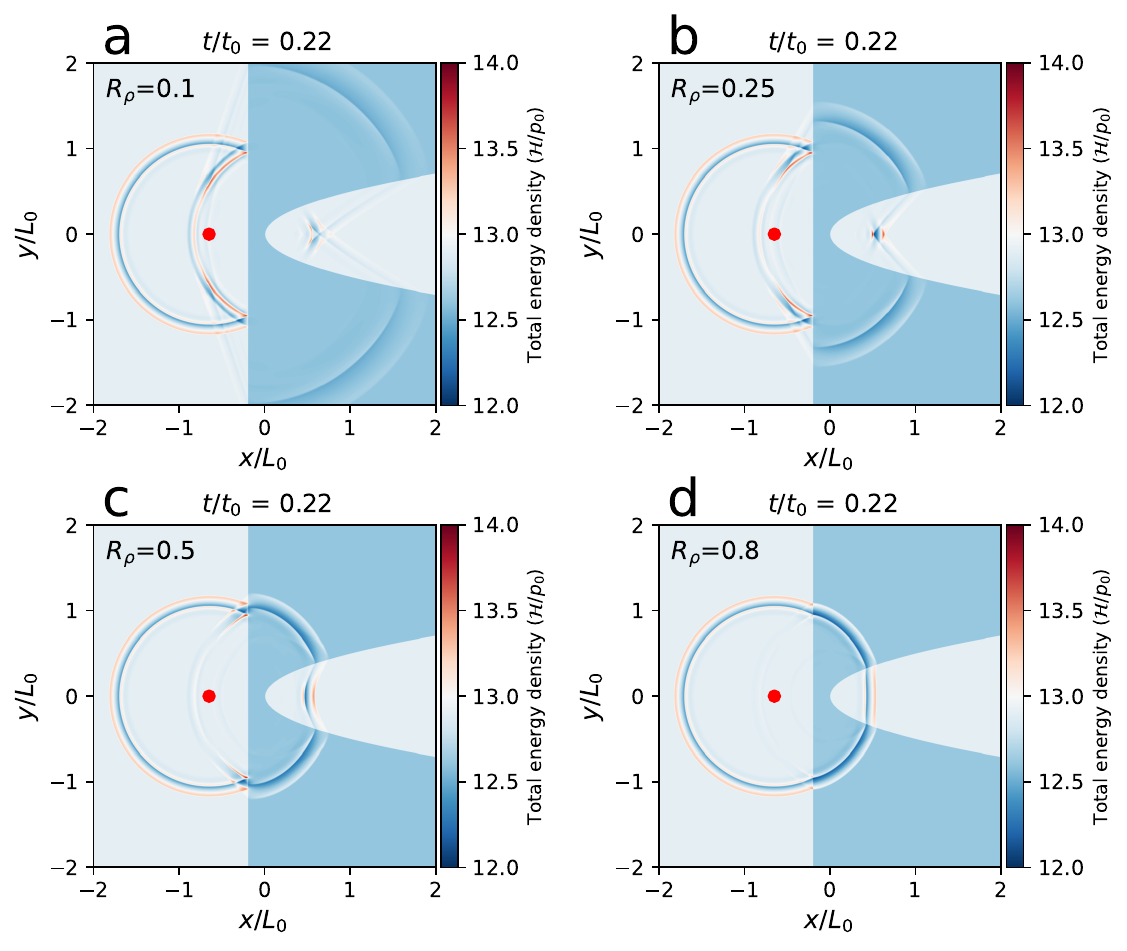}
	\caption{\textcolor{black}{{\bf MHD lensing effects with various density ratios $R_{\rho}$ between the CH and background plasma } (a)-(d) MHD lensing effect at  $R_{\rho}=0.1$, $R_{\rho}=0.25$, $R_{\rho}=0.5$, and $R_{\rho}=0.8$, respectively. The simulation domain includes a background coronal plasma (light blue) and an artificial MHD lens (blue) enclosed by a straight line and a parabolic curve. The values of normalization factors $p_0$, $t_0$, $L_0$ for the total energy density($\mathcal{H}$), time ($t$) and length ($x,y,z$) are listed in Table~\ref{tab:norm}. \label{ap2fig3}} }
\end{figure}

\begin{figure}
	\centering
	\includegraphics[width=1\textwidth]{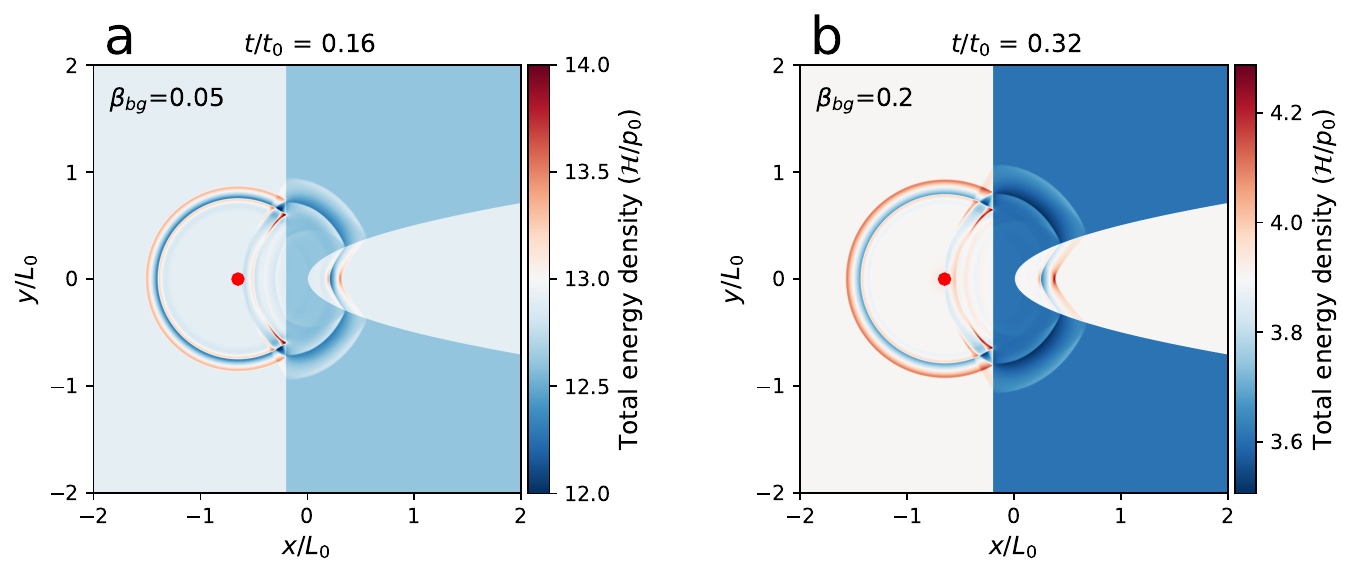}
	\caption{\textcolor{black}{{\bf MHD lensing effects with various plasma beta $\beta_{\mathrm{bg}}$ for the background coronal plasma.} (a)-(b) The MHD lensing effect when the plasma beta $\beta_{\mathrm{bg}}$ is 0.05 and 0.2. The simulation domain includes a background coronal plasma (light blue or light orange) and an artificial MHD lens (blue or deep blue) enclosed by a straight line and a parabolic curve. The values of normalization factors $p_0$, $t_0$, $L_0$ for the total energy density($\mathcal{H}$), time ($t$) and length ($x,y,z$) are listed in Table~\ref{tab:norm}. \label{ap2fig4}} }
\end{figure}

\begin{figure}
	\centering\textbf{}
	\includegraphics[width=1\textwidth]{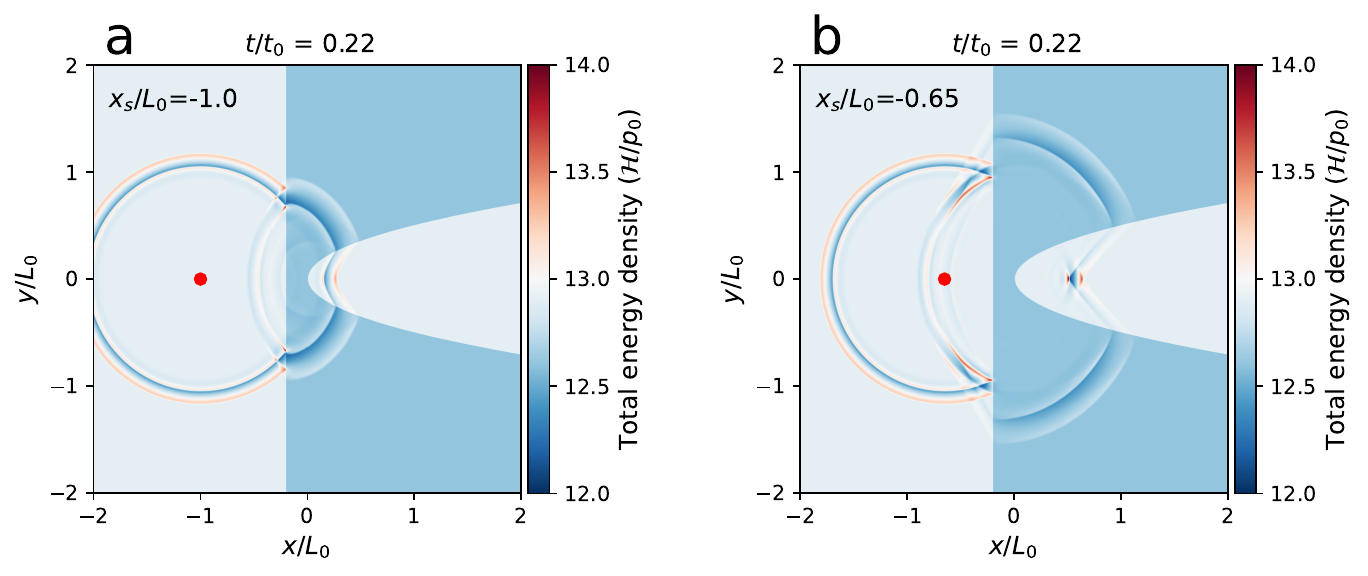}
	\caption{\textcolor{black}{{\bf MHD lensing effects with different positions $x_{s}$ of wave source.} (a)-(b) The MHD lensing effect when the relative position $x_{s}/L_0$ of the wave source is taken as -1.0 and as -0.65. The simulation domain includes a background coronal plasma (light blue) and an artificial MHD lens (blue) enclosed by a straight line and a parabolic curve. The values of normalization factors $p_0$, $t_0$, $L_0$ for the total energy density($\mathcal{H}$), time ($t$) and length ($x,y,z$) are listed in Table~\ref{tab:norm}. \label{ap2fig5}}}
\end{figure}

We also investigated the essential aspects of the MHD lensing by varying the various parameters, where the influence of the model parameters can be shown clearly. By doing so, we only adjusted one parameter in each case while keeping all other parameters unchanged. We first investigated how the MHD lensing effect changes with various density ratios $R_{\rho}$ between the CH and background plasma, as shown in \figref{ap2fig3}{}. The MHD lensing effect increases as $R_{\rho}$ decreases. When $R_{\rho}=0.1$, the transmitted wave is so strongly \textcolor{black}{bent} by the CH that a single focal point cannot be formed. The MHD lensing focal points are obvious to identify when $0.2<R_{\rho}<0.5$. When $R_{\rho}$ is increased towards $1$, the MHD lensing effect gradually disappears because the refractive indices of the CH and the background are almost identical. Then\textcolor{black}{,} we investigated how the MHD lensing effect changes with various plasma beta $\beta_{\mathrm{bg}}$ for the background coronal plasma as shown in \figref{ap2fig4}{}. The plasma beta is inversely proportional to the square of the Alfv\'en speed. Thus\textcolor{black}{,} the Alfv\'en wave propagation time is scaled with the square root of plasma beta as shown in \figref{ap2fig4}{}, where the time on the right is twice of the left and the plasma beta is 4 times. We also investigated how the position $x_{s}$ of the wave source changes the MHD lensing effect, as shown in \figref{ap2fig5}{}. The focal point of the transmitted wave shifts in the same direction to which the source position $x_{s}$ moves horizontally. 
We also discovered that changing $R_{T}$ does not have a major influence, nor does changing the amplitude $P_{b}$.

In the main paper, we present a simulation that used the observed realistic geometric shape of the CH, where the numerical setup is the same as the idealized simple setup, except for the shape of the CH.

\clearpage
\noindent{\bf Data Availability}\\
Observational data used in this study were obtained from the STEREO mission ( {\bf \color{blue}{\url{https://sdac.virtualsolar.org/cgi/search?time=1&instrument=1&version=current&build=1}}}) and SDO mission ({\bf \color{blue}{\url{http://jsoc.stanford.edu/ajax/exportdata2.html?ds=aia.lev1}}}) and are publicly available. The simulation and derived data supporting the findings of this study are available from the corresponding author on request. Source data are provided \textcolor{black}{in} this paper.

\noindent{\bf Code Availability}\\ The code of the numerical simulation is available without any restrictions \xp{on} the MPI-AMRVAC website ({\bf \color{blue}{\url{http://amrvac.org}}}). {\xp{The routines for processing observation data and PFSS model are publicly available in the solar software (SSW)} }({\bf \color{blue}{\url{http://www.lmsal.com/solarsoft/ssw_install.html}}}) .

\vspace{1.5cm}


\begin{addendum}
	\item We acknowledge the teams of STEREO and SDO for providing the excellent data. Y. Shen is supported by the National Natural Science Foundation of China (grant numbers 12173083, 11922307, 11773068), the Yunnan Science Foundation for Distinguished Young Scholars (grant number 202101AV070004), and the National Key R\&D Program of China (grant number 2019YFA0405000). D. Yuan, L. Fu and J. Wang are supported by the National Natural Science Foundation of China (grant numbers 12173012,12111530078), the Guangdong Natural Science Funds for Distinguished Young Scholar (grant number 2023B1515020049), the Shenzhen Technology Project (grant number GXWD20201230155427003-20200804151658001), and the Shenzhen Key Laboratory Launching Project (No. ZDSYS20210702140800001). R. Keppens is supported by the ERC Advanced Grant PROMINENT (funding from the European Research Council under the European Union's Horizon 2020 research and innovation programme, grant agreement No. 833251 PROMINENT ERC-ADG 2018). X. Zhao is supported by FWO project G0B4521N, and X. Zhou is supported by the National Natural Science Foundation of China (grant number 12303062) and Natural Science Foundation for Youths of Sichuan Province (grant number 2023NSFSC1351).

	\item[Author Contributions] Y.S. and D.Y. initiated the idea and led the discussion, writing, and revision of the manuscript. \textcolor{black}{X.Zhou.}, Z.T., and C.Z. contributed to \xp{the} discussion, data analysis, and manuscript writing. \textcolor{black}{X.Zhao.} and R.K. did the numerical experiment and manuscript writing. L.F. and J.W. analyzed the simulation data.
	
	\item[Competing Interests] The authors declare no competing interests.

	\clearpage

\end{addendum}


 \begin{table*}[htbp] 
	\caption{Quantities used for normalization} \label{tab:norm}
	\begin{center}
		\begin{tabular}{r r }
			\hline\hline
			Quantity & Normalization factor \\
			\hline
			Length ($x,y,z$) & $L_{0}=2.8\times10^{10}\,\mathrm{cm}$ \\
			Temperature ($T_{0}$) & $T_{0}=2\times 10^{6} \,\mathrm{K}$ \\	
			Number density ($n$) & $n_{0}=2\times 10^{8} \,\mathrm{ cm^{-3}}  $\\
			Mass density ($\rho$) & $\rho_{0}=1.4n_{0}m_{\mathrm{H}}=4.6833\times 10^{-16} \,\mathrm{g\cdot cm^{-3}}  $\\		
			Pressure ($p$) & $p_{0}=(\rho_{0} k_{B} T_{0})/(\mu  m_{\mathrm{H}}) =0.12701\,\mathrm{Ba}$ \\
			Total energy density ($\mathcal{H}$) &  $p_{0}=0.12701 \, \mathrm{Ba}$ \\			
			Magnetic field strength ($\mathbf{B}$) & $B_{0}=\sqrt{4\pi p_{0}}=1.2634\,\mathrm{G}$ \\	
			Velocity ($v_{0}$) & $v_{0}=B_{0}/\sqrt{4\pi \rho_{0}} =1.6468\times 10^{7} \,\mathrm{cm\cdot s^{-1}}$\\		
			Time ($t$)&  $t_{0}=L_{0}/v_{0}=1700.2\,\mathrm{s}$ \\ 
			\hline
		\end{tabular}  
	\end{center}
	\label{table}
\end{table*}

\end{document}